Deep Reinforcement Learning Designed RadioFrequency Waveform in MRI

Dongmyung Shin, Younghoon Kim, Chungseok Oh, Hongjun An, Juhyung Park, Jiye Kim, and Jongho Lee

*Department of Electrical and Computer Engineering, Seoul National University, Seoul, Korea*

*Correspondence should be addressed to Jongho Lee (jonghoyi@snu.ac.kr)*

## ABSTRACT

Carefully engineered radiofrequency (RF) pulses play a key role in a number of systems such as mobile phone, radar, and magnetic resonance imaging. The design of an RF waveform, however, is often posed as an inverse problem with no general solution. As a result, various design methods each with a specific purpose have been developed based on the intuition of human experts. In this work, we propose an artificial intelligence (AI)-powered RF pulse design framework, DeepRF, which utilizes the self-learning characteristics of deep reinforcement learning to generate a novel RF pulse. The effectiveness of DeepRF is demonstrated using four types of RF pulses that are commonly used. The DeepRF-designed pulses successfully satisfy the design criteria while reporting reduced energy. Analyses demonstrate the pulses utilize new mechanisms of magnetization manipulation, suggesting the potentials of DeepRF in discovering unseen design dimensions beyond human intuition. This work may lay the foundation for an emerging field of AI-driven RF waveform design.

In less than a decade, artificial intelligence (AI) has been widely applied to a number of tasks which were once considered as unique human abilities, such as winning the game Go[1], translating books[2], and even blushing paints[3]. More recently, AI-driven scientific designs have become a new direction for research, finding solutions in chemical synthesis[4], RNA design[5], electric circuit design[6], and drug discovery[7]. In particular, deep reinforcement learning (DRL), in which an AI agent learns a skill by self-exploring an environment and receiving feedback (or reward) from the environment[8], has been utilized to achieve novel design results[5-7] that surpass conventional outcomes by discovering unseen opportunities that exist beyond human intuition.

A radiofrequency (RF) pulse plays an important role in numerous applications, including broadcasting (e.g., radio and TV), wireless technology (e.g., mobile phone, satellite, and WiFi), military (e.g., radar), and medical systems (e.g., ultrasound, RF ablation, and magnetic resonance imaging). In many of these areas, the careful design of an RF pulse shape is required to achieve a complex goal. For example, in a radar system, an RF pulse is created to classify target objects (e.g., air balloon vs. fighter jet)[9]. In ultrasound imaging, a set of RF pulses are jointly designed to image microbubbles in tissue[10]. In magnetic resonance imaging (MRI), a variety of RF pulses are developed to control tissue magnetization, forming images.

In a broad sense, such RF pulse designs are subject to an inverse problem[11], which is a process of calculating causal factors from a set of observations in a given model. In MRI, for example, an RF pulse (i.e., causal factor) is designed to achieve desired tissue magnetization (i.e., observation) according to Bloch equations[12] (i.e., model). This task, however, has no general solution. As a result, various design methods each with a specific purpose have been developed based on the intuition of human experts (e.g., Shinnar-Le Roux (SLR) algorithm for a slice-selective RF pulse[13] and adiabatic principle for a $B_1$-insensitive RF pulse[14]) or optimization algorithms (e.g., optimal control[15-22] and convex optimization[23,24]).

In this study, we propose a new RF pulse design framework, DeepRF (<u>Deep</u> Reinforcement Learning Designed <u>RF</u> Waveform), which utilizes the power of the self-learning characteristic of DRL, combined with optimization, to explore novel mechanisms for magnetization manipulation beyond conventional methods. This feature clearly differentiates DeepRF from previously proposed deep learning-based RF design methods using supervised learning[25-28]. Moreover, unlike conventional RF design algorithms that are tailored for a specific RF pulse type, DeepRF can produce RF pulses of different goals via a customized reward function. Here, we develop four DeepRF pulses of different characteristics, slice-selective excitation pulse, slice-selective inversion pulse, $B_1$-insensitive volume inversion pulse, and $B_1$-insensitive selective inversion pulse, that are commonly utilized in MRI. Additionally, analyses on the characteristics of DeepRF pulses, reproducibility, and ablation studies of the DeepRF algorithm are included to explore the newly proposed method.

**DeepRF algorithm**

An overview of DeepRF is summarized in Fig. 1. DeepRF is performed as a sequential execution of two modules: RF generation and RF refinement (Fig. 1a). In the RF generation module, DRL produces a large number of RF pulses, some of which are utilized as seed RF pulses for the RF refinement module, which further improves the characteristics of the pulses (Fig. 1a). The DRL in the RF generation module utilizes a recurrent neural network (RNN; see **RNN agent** in **Supplementary Information**)[29] agent to produce an RF pulse (i.e., actions), which is evaluated by a virtual MRI simulation (i.e., environment), calculating a reward (e.g., difference from a target slice profile) (Fig. 1b). This reward is fed back to the agent, updating weights of RNN for improved RF pulse generation. Through this iteration, pulses with higher rewards are produced. Finally, the top 256 RFs out of 38,400,000 RFs are selected as the seed RFs for the RF refinement module (see **RF generation module** in **METHODS** for details).

In the RF refinement module, the 256 RFs are iteratively refined via gradient ascent (Fig. 1c). For the RF pulses of $i^{th}$ gradient ascent iteration, the virtual MRI simulation generates slice/inversion profiles, calculating rewards for each RF. The derivatives of the average reward, $Reward_{avg}$, are calculated, then scaled, and added to the $i^{th}$ RFs, creating the next $(i+1)^{th}$ RFs. After 10,000 iterations, the RF pulse with the highest reward is chosen as the final result (see **RF refinement module** in **METHODS** for details).

To demonstrate that DeepRF generates effective RF pulses of different goals, we design four types of RF pulses that have matching characteristics with commonly used conventional RF pulses: slice-selective excitation pulse and slice-selective inversion pulse using the SLR algorithm[13]; $B_1$-insensitive volume inversion pulse and $B_1$-insensitive selective inversion pulse using the hyperbolic secant (HS) pulse design[30] (see **Conventional pulses** in **METHODS**). All pulses are designed from scratch. A reward function is customized for each of the RF pulses (see **Reward functions** in **METHODS**). All DeepRF and conventional RF pulses are evaluated by simulations and experiments for the characteristics of pulse shapes such as slice profiles and RF energy (ENG; $\int_0^T |B_1(t)|^2 dt$ where $T$ is the pulse duration, $B_1$ is the pulse shape, and unit is Gauss$^2$·second; see **METHODS**). The computational times of the DeepRF and conventional pulses are reported. Note that the computational time of a DeepRF pulse is expected to be high because of the large number of pulses generated and optimized via iterations to ensure high reproducibility (see **Computational time of DeepRF** in **Supplementary Information**).

**Slice-selective excitation pulse**

To design a slice selective excitation pulse using DeepRF, we define the reward function to have the same mean transverse magnitude over a bandwidth (BW; full width at half maximum) and the same maximum stopband ripple (= 1.5%) as those of an SLR excitation pulse (parameters: angle = 90°, duration = 2.56 ms, time-bandwidth product (TBW) = 6.6, least-squares design; see **Conventional pulses** in **METHODS**) while minimizing RF ENG. The results show a reduced ENG while retaining a comparable slice profile to the conventional RF design (Fig. 2a-h). The amplitude of the DeepRF pulse mimics that of the SLR pulse (Fig. 2a) while showing a lower peak amplitude, which results in 17% smaller ENG in the DeepRF pulse. The phase of the DeepRF pulse is also similar but reveals a continuous change compared to that of the SLR pulse (Fig. 2e). The slice profiles of the computer simulation are close to each other (Fig. 2b; zoomed-in image in Fig. 2f), reporting the same mean transverse magnitude over the BW (= 0.93 in Fig. 2f). The experimental slice profiles match the simulation results well (Fig. 2b vs. Fig. 2c; Fig. 2f vs. Fig. 2g) and also report similar mean transverse magnitude over the BW for both designs (= 0.93 for SLR and 0.92 for DeepRF in Fig. 2g). The maximum stopband ripples (= 1.5%) are the same (Fig. 2h), while the maximum passband ripple, which is not a design criterium in DeepRF, is higher in the DeepRF pulse (2.2% for DeepRF and 0.4% for SLR in Fig. 2d; see **Alternative designs** and **Maximum passband ripple of slice-selective excitation pulse** in **Supplementary Information**). The computational time for DeepRF was approximately 40 hours whereas that of SLR was less than a second. When the DeepRF results are compared with those of the equi-ripple-designed SLR pulse, ENG gain is slightly improved (= 20%), and the slice profiles are consistent (mean transverse magnitude = 0.92; maximum stopband ripple = 1.2%; see Supplementary Fig. 1). To demonstrate the scalability of DeepRF for a design parameter, we perform additional simulations with different TBWs (8.2, 9.9, and 11.5), and the results are summarized in Supplementary Fig. 2, reporting consistent ENG efficiency.

### Slice-selective inversion pulse

The summary of the slice-selective inversion pulse design is shown in Fig. 2i-p. The pulses are designed with a pulse duration of 5.12 ms, TBW of 4.3, and least-squares design (see **Conventional pulses** in **METHODS** for details). For DeepRF, the reward function is defined to reduce the mean squared error (MSE) of z-magnetization ($M_z$) between DeepRF and SLR profiles while minimizing RF ENG. Unlike the excitation pulse, both amplitude and phase shapes of the DeepRF pulse are substantially different from those of the SLR pulse (Fig. 2i and m). This suggests that a different inversion mechanism may exist in the DeepRF pulse, which is illustrated later (see **Analyses of DeepRF pulses**). The ENG of the DeepRF pulse is smaller by 11% than that of the SLR pulse (Fig. 2i). The simulated slice profiles are highly comparable (Fig. 2j and n), showing the same values of the mean longitudinal magnitude over the BW (mean $M_z$ over a BW; -0.81 in Fig. 2n). The experimental results also show good matches with the simulation results (Fig. 2j vs. Fig. 2k; Fig. 2n vs. Fig. 2o) and report the same mean $M_z$ over the BW (= -0.76 in Fig. 2o). The maximum stopband ripples are the same in both designs (0.3% in Fig. 2p), whereas the maximum passband ripple is slightly higher in the DeepRF profile (1.9% for DeepRF and 1.7% for SLR; not included in the design criteria for DeepRF). The computational times were the same as the excitation pulses. Similar findings are observed when the DeepRF results are compared with those of the equi-ripple-designed SLR pulse (the same ENG gain, mean $M_z$ over a BW, and maximum stopband ripple; see Supplementary Fig. 3). When three additional results with different TBWs (5.6, 6.8, and 8.1) are tested for scalability, the results reveal consistent slice profiles to the conventional method while providing ENG efficiency (Supplementary Fig. 4).

**$B_1$-insensitive volume inversion pulse**

The results of the $B_1$-insensitive volume inversion pulse design are shown in Fig. 3a-h. The conventional HS pulse is optimized to achieve the mean $M_z$ of -0.9 over the target region (frequency range = -200 Hz to 200 Hz, $B_1$ range = 0.5 to 2.0; see **Conventional pulses** in **METHODS**). The reward function of DeepRF is defined to satisfy the same mean $M_z$ over the target region while minimizing RF ENG. When the amplitude shape of the DeepRF pulse is compared with that of the HS pulse, it shows a substantially different characteristic (Fig. 3a). The ENG of the DeepRF pulse is smaller by 9% than that of the HS pulse, demonstrating the advantage of the new method. The simulated inversion profiles (i.e., $M_z$ for a range of $B_1$ and frequencies; Fig. 3b and f) reveal that the mean $M_z$ over the target region (red dotted boxes) are the same (= -0.906 for both RFs), although the details of the profiles in Fig. 3d and h are different. For the profiles outside of the target region, no constraint was imposed, and the results reveal the substantial difference between the two pulses. The experimental profiles closely mimic the simulated profiles (Fig. 3b vs. Fig. 3c; Fig. 3f vs. Fig. 3g), reporting similar values of the mean $M_z$ over the target region (-0.886 for HS and -0.890 for DeepRF). The computational time for DeepRF was approximately 27 hours whereas that of HS using a simple grid search was 5 hours. When three additional DeepRF and HS pulses with different target regions (frequency range = -300 Hz to 300 Hz, -400 Hz to 400 Hz, and -500 Hz to 500 Hz) are produced, they report consistent ENG reductions in the DeepRF pulses (Supplementary Fig. 5).

**$B_1$-insensitive selective inversion pulse**

For the $B_1$-insensitive selective inversion pulses, the reward function is defined to reduce the MSE of $M_z$ between the DeepRF and HS profiles while minimizing RF ENG. The HS pulse is designed to have the mean $M_z$ of -0.9 over the target region (frequency range = -200 Hz to 200 Hz, $B_1$ range = 0.5 to 2.0; see **Conventional pulses** in **METHODS**). The results are shown in Fig. 3i-p. The amplitude and phase shapes of the two pulses are surprisingly similar to each other (Fig. 3i and m), despite the fact that no knowledge of the adiabatic principle[14] nor any of the conventional RF pulses are utilized for training DeepRF. The DeepRF pulse results in a small ENG reduction (= 2%). The simulated profiles from both pulses are well-matched (Fig. 3j vs. Fig. 3n), and the experimental profiles agree well with the simulated profiles (Fig. 3j vs. Fig. 3k; Fig. 3n vs. Fig. 3o). The zoomed-in versions of the experimental profiles report the mean $M_z$ over the target region of -0.886 for HS and -0.892 for DeepRF (red dotted boxes in Fig. 3l and p). The computational time for DeepRF was approximately 56 hours whereas that of HS using a grid search was 5 hours. When simulated for three additional RFs with different target regions (frequency range = -300 Hz to 300 Hz, -400 Hz to 400 Hz, and -500 Hz to 500 Hz), the results show matching pulse shapes and profiles (Supplementary Fig. 6).

**Analyses of DeepRF pulses**

To understand the characteristics of the DeepRF pulses, we explore the temporal changes of the magnetizations in the slice-selective inversion pulse and $B_1$-insensitive volume inversion pulse that show different RF shapes from the conventional pulses. For the slice-selective inversion pulse, the progression of the slice profile is plotted at six-time points of the RF pulse (from 1.92 ms to 5.12 ms; Fig. 4a-f). The results demonstrate that the DeepRF pulse may have a different inversion process from that of the SLR pulse.

For the $B_1$-insensitive volume inversion pulse, the analysis is performed for two different conditions of $B_1$ and off-resonance frequencies ($B_1$ = 1.0 and frequency = 0 Hz; $B_1$ = 1.5 and frequency = 150 Hz). When the trajectory of the spin is plotted over the pulse duration (Fig. 4g vs. Fig. 4k; Fig. 4i vs. Fig. 4m), the DeepRF and HS pulses reveal substantially different trajectories, suggesting different mechanisms for inversion. For further analysis, we examine the adiabaticity of the two pulses ($K(t) = |\frac{\omega_{eff}(t)}{\dot{\alpha}}|$ where $\omega_{eff}$ is an effective magnetic field and $\dot{\alpha}$ is an instant angular velocity of $\omega_{eff}$)[14,31,32] (Fig. 4h, j, l, and n). In the HS pulse, the minimum adiabaticity is 3.6 and 3.1 for the two conditions (Fig. 4h and j), satisfying the adiabatic condition throughout the pulse duration (i.e., $K(t) \gg 1$). In the DeepRF pulse, on the other hand, adiabaticity is often smaller than the minimums of the HS pulse (red dotted lines in 4h, j, l, and n), reporting the minimum adiabaticity of 0.1 in Fig. 4l and 0.04 in Fig. 4n. These results suggest that DeepRF severely violates the conventional adiabatic condition and indicate that the DeepRF pulse utilizes an inversion mechanism that is difficult to be explained by the conventional design principle.

**Reproducibility**

Because DRL is known for low reproducibility due to stochasticity in the learning process[33], we explore the reproducibility of DeepRF by repeating each design three times using different initialization via Xavier[34]. The results demonstrate that the RF shapes and slice/inversion profiles of the repetitions are similar (Supplementary Figs. 7-10), reporting consistent RF ENG reductions and slice/inversion profile metrics (Table 1; all values are given as mean ± standard deviation). In particular, the two $B_1$-insensitive inversion pulse designs generate almost identical pulse shapes over the repetitions (Supplementary Figs. 9 and 10).

**Ablation study**

To examine the effects of the RF generation module or the RF refinement module on the final RF pulse, we ablate each module in the design process of DeepRF (Fig. 1a). When the RF generation module is removed, and the four types of RFs are repeatedly designed three times with random seeds (amplitude: uniform sampling from 0 to 0.2 Gauss; phase: from -π to π), the results are failures, reporting no reduction in RF ENG or lower ENG gains with poor reproducibility, except for the $B_1$-insensitive volume inversion pulse (Supplementary Table 1 and Supplementary Figs. 11-14; see **Ablation study of $B_1$-insensitive volume inversion pulse** in **Supplementary Information**). In particular, the ablation results of the $B_1$-insensitive selective inversion pulse suggest that DeepRF without DRL is incapable of designing an RF pulse that satisfies the target inversion profile (Supplementary Fig. 14).

When the RF refinement module is ablated, the slice/inversion profiles and/or RF ENG reductions are poorer than those of the original DeepRF pulses, demonstrating the needs for both modules (Supplementary Figs. 15-18). For the slice-selective excitation and inversion pulses, the ablation results reveal no reduction in ENG and degraded slice profiles (Supplementary Figs. 15 and 16). The ablation results of the $B_1$-insensitive volume inversion pulse report no reduction in ENG (Supplementary Fig. 17), and the results of the $B_1$-insensitive selective inversion pulse show a degraded inversion profile (Supplementary Fig. 18).

**Alternative designs**

In this work, the reward functions are defined to match the design specifications of the conventional RF pulses (matching design spec; slice-selective excitation RF and $B_1$-insensitive volume inversion RF) or to minimize the MSE between the inversion profiles of the two pulses (minimizing MSE; slice-selective inversion RF and $B_1$-insensitive selective inversion RF). For additional investigations, we test alternative reward functions (see **Alternative reward functions** in **METHODS**). Supplementary Fig. 19 summarizes the alternative design (minimizing MSE) results of the slice-selective excitation pulse, revealing a lower ENG gain than that of the original DeepRF design (5% for minimizing MSE vs. 17% for matching design spec (original)). For the $B_1$-insensitive volume inversion pulse, the results also report a lower ENG gain than that of the original DeepRF design (4% for minimizing MSE vs. 9% for matching design spec (original); Supplementary Fig. 20). When we tested the reward functions of matching design specifications for the slice-selective inversion RF and $B_1$-insensitive selective inversion RF, the results show the same or improved ENG gain with slightly different slice profiles (see Supplementary Figs. 21 and 22 for details).

**DISCUSSION**

In this study, we proposed an AI-powered RF design method and demonstrated the method could successfully design four commonly used RF pulses in MRI. In particular, the power of self-learning characteristics in DRL, which was well-highlighted in Go[1] and StarCraft[35], enabled the method to discover novel RF pulses, revealing new mechanisms of magnetization manipulation beyond conventional examples. Furthermore, the sequential combination of the RF generation module, which acted as an exploration step, searching for new RF pulse candidates, and the RF refinement module, which served as an exploitation step, optimizing the seed RF pulses, substantially improved the final outcome as demonstrated in the ablation study. Finally, the reproducibility results support the stability of our algorithm, yielding reliability for future applications.

Recently, supervised learning-based RF pulse designs have been demonstrated to create multidimensional or parallel transmit RF pulses, providing advantages in inference time (7 ms to 130 ms)[25-28]. Different from these studies, DeepRF can produce various types of RF pulses as shown in this work. Moreover, DeepRF performs self-learning, allowing the network to explore new strategies as demonstrated in the example of AlphaGo Zero[1].

In previous studies of MRI, DRL was suggested to develop a part of a sequence, such as encoding gradient[36], sequence parameters[37], and k-space sampling trajectory[38-40]. For example, Bo Zhu, et al. demonstrated DRL could generate a readout gradient that mimicked conventional Fourier encoding[36]. In the work by Samuel, et al., DRL was utilized for the real-time control of pulse sequence parameters (e.g., repetition time and flip angle) to guess the shape of a scanned object[37]. In the studies by Zeng, et al. and Pineda, et al., k-space under-sampling patterns were optimized with respect to reconstructed image quality[38,39]. While these methods applied DRL for designing a part of a pulse sequence, none of them designed novel RF pulses nor investigated their mechanisms, differentiating our work from them (see **DeepRF$_{SLR}$** in **Supplementary Information**).

The RF refinement module shares similarity to optimal control (OC)-based RF pulse design methods[15-22] which update an RF pulse iteratively via a numerical optimization method to maximize (or minimize) an objective function. When we implemented the optimal control method with energy-constraint[15] and compared the results with those of DeepRF and SLR for the slice-selective excitation and inversion pulses, DeepRF pulses showed larger ENG reduction than the OC pulses (OC: -4% for excitation and -4% for inversion; DeepRF: -17% for excitation and -11% for inversion; Fig. 5), although the reduction may be improved via optimization.

In this study, the feasibility of designing a DeepRF pulse was demonstrated for commonly utilized RF pulses. The method, however, is not restricted to these pulses and may be applied to develop more complicated pulses such as a spatial-spectral pulse[41], a multidimensional pulse[42], etc. Furthermore, a more complex-shaped slice profile that has not been suggested might be created via a careful design of the reward function. Additionally, it may be utilized to design gradient waveforms as well, potentially opening a new way toward designing a new sequence.

So far, we have restricted our attention to an RF pulse design in MRI, where the physical model is governed by Bloch equations[12]. One might expand the concept of DeepRF to design waveforms of other systems that have different physical models such as the point target model for radar[9] or scattering model for ultrasound[10]. We expect DeepRF may be utilized in such systems to design a more effective RF waveform of a new mechanism, as demonstrated in this study.

**METHODS**

**RF generation module**

In the RF generation module, a large number of RF pulses (38,400,000 RFs) are produced using DRL (Fig. 1b) to ensure reproducibility. DRL consists of the following components. The agent is an RNN[29]. The state represents a pair of amplitude and phase values at each RF time step (in radians; see Supplementary Table 2 for hyperparameters). The action is the generation of this pair. The environment is a virtual MRI simulation, which produces a slice/inversion profile for an input RF. The reward function is customized for each type of RF pulse (e.g., reduce RF ENG while matching the target slice profile; see **Reward functions**).

The RNN consists of a gated recurrent unit (GRU)[43] and four cascaded fully-connected (FC) layers (Supplementary Fig. 23a and 24). At the time step t, the GRU takes an amplitude and phase RF pair $(A_t, P_t)$ and a hidden state vector ($\mathbf{H_t} \in R^{256}$), which carries the information of RF pairs in previous time steps with $\mathbf{H_0}$ being a zero vector, as the input and produces a next hidden state as the output. This output is fed to the FC layers, generating the mean values for the normal distributions of the amplitude ($\mu_{A,t+1}$) and phase ($\mu_{P,t+1}$), and value estimate ($V_{t+1}$) which is used to stabilize the learning process[44]. Finally, the amplitude and phase values of the next time step $(A_{t+1}, P_{t+1})$ are sampled from the two normal distributions, $N(\mu_{A,t+1}, \sigma_A^2)$ and $N(\mu_{P,t+1}, \sigma_P^2)$ where $\sigma_A^2$ and $\sigma_P^2$ are fixed variance values (see Supplementary Table 2).

The detailed procedure of generating an RF pulse is summarized in Supplementary Fig. 23b. The RNN agent sequentially creates T (= 32) number of amplitudes, phases, and value estimates (i.e., $[(A_1, P_1, V_1) \dots (A_T, P_T, V_T)]$). After that, the amplitudes and phases are mirrored and concatenated, producing $[(A_1, P_1) \dots (A_T, P_T)(A_T, P_T) \dots (A_1, P_1)]$, which is then up-sampled to $[(A_1^u, P_1^u) \ ... \ (A_L^u, P_L^u)]$ using linear interpolation (L = 256), completing an RF generation. The mirroring and interpolation procedures are motivated by the observation of symmetric and slow-varying nature in conventional RF pulses (e.g., SLR[13] and HS[30] pulses), although asymmetric designs exist (e.g., minimum phase SLR inversion pulse). The final RF pulse is fed into the virtual MRI simulation, generating a slice/inversion profile. Then, a reward is evaluated using this profile and the ENG of the RF pulse. All the data of the RF pulse (i.e., amplitudes, phases, value estimates, and reward) are saved into a buffer.

When the buffer is filled up with a prefixed number (= 256) of the RF pulse data, RNN weights are updated using the proximal policy optimization method[44] (see **Loss function of DRL** in **Supplementary Information**). A single run of the RF generation module updates the RNN weights 300 times, generating 768,000 RF pulses. The run is repeated 50 times with different initialization of the RNN, creating 38,400,000 RF pulses in the RF generation module. Among them, the top 256 RF pulses with the highest rewards are selected as seed RFs for the RF refinement module. The number of the seed RFs is limited by GPU memory.

**RF refinement module**

The final RF pulses of the RF generation module are utilized as seed RFs ($[\mathbf{RF}_1^0 \ \cdots \ \mathbf{RF}_{256}^0] \in \mathbb{C}^{L \times 256}$) for the RF refinement module. These RF pulses are iteratively refined using gradient ascent (Fig. 1c). For the RF pulses of i$^{th}$ iteration ($[\mathbf{RF}_1^i \ \cdots \ \mathbf{RF}_{256}^i] \in \mathbb{C}^{L \times 256}$), the virtual MRI simulation generates 256 slice/inversion profiles. Then, these profiles and ENGs of the RF pulses are used to calculate 256 reward values (see **Reward functions**), which are averaged for an averaged reward function, $Reward_{avg}$. Then the derivatives of $Reward_{avg}$ with respect to the input ($\left[\frac{\partial Reward_{avg}}{\partial RF_1^i} \cdots \frac{\partial Reward_{avg}}{\partial RF_{256}^i}\right] \in \mathbb{C}^{L \times 256}$) are calculated using reverse-mode automatic differentiation[45], finding derivatives for individually pulses. Finally, the RF pulses are updated as follows:

$$\left[RF_1^{i+1} \cdots RF_{256}^{i+1}\right] = \left[RF_1^i \cdots RF_{256}^i\right] + \alpha \cdot \left[\frac{\partial Reward_{avg}}{\partial RF_1^i} \cdots \frac{\partial Reward_{avg}}{\partial RF_{256}^i}\right] \quad (1)$$

where $\alpha$ is a step size of the gradient ascent using Adam optimizer[46] with the initial input step size of 0.01, determined empirically. The gradient ascent iteration is repeated 10,000 times. After the iterations, the RF pulse with the highest reward is chosen as the final design result of DeepRF.

**Reward functions**

DeepRF is a goal-oriented RF design framework and, therefore, can produce RF pulses of various characteristics via constraints in the reward function. In our study, two types of reward functions are defined: One is to satisfy the design specifications of a conventional RF pulse, and the other is to reduce the difference in the slice/inversion profile from that of a conventional RF pulse. A detailed description of the reward function for each of the four RF pulses is as follows.

The reward function of the slice-selective excitation pulse (Fig. 2) is designed to have the same mean transverse magnitude over a BW (full width at half maximum) and the same maximum stopband ripple as those of the SLR slice profile while minimizing ENG. This is defined as follows:

$$\min\left(\sqrt{E[M_{x,DeepRF}(f_b)]^2 + E[M_{y,DeepRF}(f_b)]^2}, c_1\right) -$$

$$\max\left(\max_{element}\left(\sqrt{M_{x,DeepRF}(f_s)^2 + M_{y,DeepRF}(f_s)^2}\right) - c_2, 0\right) - c_3 \cdot ENG_{RF} \quad (2)$$

where $E[\ ]$ is an expectation, $M_{x,DeepRF}(f)$ and $M_{y,DeepRF}(f)$ are x- and y-magnetizations over a frequency range ($f$), $f_b$ and $f_s$ represent BW and stopband frequency ranges ($f_b$: from -1285 Hz to 1285 Hz with a step size of 2.6 Hz; $f_s$: from -32000 Hz to -1614 Hz and 1614 Hz to 32000 Hz with a step size of 20.3 Hz), respectively, $c_1$, $c_2$ and $c_3$ are constants, and $ENG_{RF}$ represents the ENG of the RF pulse in $\mu G^2 sec$. In the first term, $\sqrt{E[M_{x,DeepRF}(f_b)]^2 + E[M_{y,DeepRF}(f_b)]^2}$ denotes the mean transverse magnitude over the BW, which is bounded by $c_1$ (= 0.92724; the mean transverse magnitude over the BW of the SLR profile). In the second term, $\sqrt{M_{x,DeepRF}(f_s)^2 + M_{y,DeepRF}(f_s)^2}$ represents the magnitudes at the stopband (i.e., stopband ripples), and the second term returns zero if the maximum stopband ripple is smaller than $c_2$ (= 0.0146) which is the maximum stopband ripple of the SLR profile. Finally, the last term penalizes the ENG of the RF pulse ($c_3 = 10^{-5}$). These constants and all the subsequent constants along with other hyperparameters (see Supplementary Table 2) were empirically determined.

For the slice-selective inversion pulse (Fig. 2), the reward function is set to reduce the mean squared error (MSE) of $M_z$ between the DeepRF and SLR profiles while minimizing ENG, which is defined as follows:

$$-c_1 \cdot MSE\left(M_{z,DeepRF}(f), M_{z,SLR}(f)\right) - c_2 \cdot ENG_{RF} \quad (3)$$

where $f$ is a frequency range from -8000 Hz to 8000 Hz with step sizes of 1.2 Hz within the BW and 5.0 Hz outsides of the BW, $\mathbf{M_{z,DeepRF}}$ and $\mathbf{M_{z,SLR}}$ are the longitudinal magnetizations of the DeepRF and SLR profiles, respectively, $c_1$ (= 0.25) and $c_2$ (= $10^{-6}$) are constants.

In the case of the $B_1$-insensitive volume inversion pulse (Fig. 3), the reward function is designed to achieve the mean $M_z$ of -0.9 over the target region (frequency range = -200 Hz to 200 Hz and $B_1$ range = 0.5 to 2.0) while minimizing ENG. This is defined as follows:

$$-\max\left(E[\mathbf{M_{z,DeepRF}}(\mathbf{B_1}, \mathbf{f})], c_1\right) - c_2 \cdot \text{ENG}_{RF} \qquad (4)$$

where $\mathbf{B_1}$ is $B_1$ scaling factors from 0.5 to 2.0 with a step size of 0.03, $\mathbf{f}$ is a frequency range from -200 Hz to 200 Hz with a step size of 8 Hz, and $c_1$ (= -0.9; target mean $M_z$) and $c_2$ (= 0.004) are constants. The maximum value of $E[\mathbf{M_{z,DeepRF}}(\mathbf{B_1}, \mathbf{f})]$ is bounded by $c_1$.

The reward function of the $B_1$-insensitive selective inversion pulse (Fig. 3) is set to reduce the MSE of $M_z$ between the DeepRF and HS profiles while minimizing ENG. The definition is as follows:

$$-c_1 \cdot \text{MSE}\left(\mathbf{M_{z,DeepRF}}(\mathbf{B_1}, \mathbf{f}), \mathbf{M_{z,HS}}(\mathbf{B_1}, \mathbf{f})\right) - c_2 \cdot \text{ENG}_{RF} \qquad (5)$$

where $\mathbf{B_1}$ is $B_1$ scaling factors from 0.5 to 2.0 with a step size of 0.03, $\mathbf{f}$ is a frequency range from -8000 Hz to 8000 Hz with step sizes of 10 Hz within the target region (-200 Hz to 200 Hz) and 75 Hz outside of the target region, $\mathbf{M_{z,DeepRF}}$ and $\mathbf{M_{z,HS}}$ are the longitudinal magnetizations of the DeepRF and HS profiles, respectively, and $c_1$ (= 0.25) and $c_2$ (= $10^{-5}$) are constants.

**Alternative reward functions**

For all pulses, additional RF pulse designs (Supplementary Figs. 19 to 22) were performed using alternative reward functions different from those of the original designs (Figs. 2 and 3). The alternative reward functions are defined as follows.

The alternative reward function of the slice-selective excitation pulse (Supplementary Fig. 19) is set to reduce the sum of the MSEs of x- and y-magnetizations between the DeepRF and SLR profiles while minimizing ENG, which is defined as follows:

$$-c_1 \cdot \left( \text{MSE}\left(\mathbf{M}_{x,\text{DeepRF}}(\mathbf{f}), \mathbf{M}_{x,\text{SLR}}(\mathbf{f})\right) + \text{MSE}\left(\mathbf{M}_{y,\text{DeepRF}}(\mathbf{f}), \mathbf{M}_{y,\text{SLR}}(\mathbf{f})\right) \right) - c_2 \cdot \text{ENG}_{\text{RF}}$$

(6)

where $\mathbf{f}$ is a frequency range from -32000 Hz to 32000 Hz with step sizes of 3.1 Hz within the BW (-1285 Hz to 1285 Hz) and 20.3 Hz outside of the BW, $\mathbf{M}_{x,\text{DeepRF}}$ and $\mathbf{M}_{y,\text{DeepRF}}$ are the x- and y-magnetizations of the DeepRF profile, respectively, $\mathbf{M}_{x,\text{SLR}}$ and $\mathbf{M}_{y,\text{SLR}}$ are the x- and y-magnetizations of the SLR profile, respectively, and $c_1$ (= 0.125) and $c_2$ (= $10^{-6}$) are constants.

The alternative reward function of the $B_1$-insensitive volume inversion pulse (Supplementary Fig. 20) is designed to reduce the MSE of $M_z$ between the DeepRF and HS profiles while minimizing ENG. The definition is as follows:

$$-c_1 \cdot \text{MSE}\left(\mathbf{M}_{z,\text{DeepRF}}(\mathbf{B_1}, \mathbf{f}), \mathbf{M}_{z,\text{HS}}(\mathbf{B_1}, \mathbf{f})\right) - c_2 \cdot \text{ENG}_{\text{RF}} \quad (7)$$

where $\mathbf{B_1}$ is $B_1$ scaling factors from 0.5 to 2.0 with a step size of 0.03, $\mathbf{f}$ is a frequency range from -200 Hz to 200 Hz with a step size of 8 Hz, $\mathbf{M}_{z,\text{DeepRF}}$ and $\mathbf{M}_{z,\text{HS}}$ are the longitudinal magnetizations of the DeepRF and HS profiles, respectively, and $c_1$ (= 0.25) and $c_2$ (= $10^{-5}$) are constants.

The alternative reward function of the slice-selective inversion pulse (Supplementary Fig. 21) is defined to match the design specifications (the same mean $M_z$ over BW and the same maximum stopband ripple as those of the SLR inversion pulse) while minimizing ENG. The definition is given as follows:

$$- \max(E[\mathbf{M_{z,DeepRF}}(\mathbf{f_b})], c_1) - \max\left(\max_{\text{element}}(|1 - \mathbf{M_{z,DeepRF}}(\mathbf{f_s})|) - c_2, 0\right) - c_3 \cdot \text{ENG}_{\text{RF}}$$

(8)

where E[ ] is an expectation, $\mathbf{M_{z,DeepRF}}(\mathbf{f})$ is z-magnetizations over a frequency range ($\mathbf{f}$), $\mathbf{f_b}$ and $\mathbf{f_s}$ represent BW and stopband frequency ranges ($\mathbf{f_b}$: from -418 Hz to 418 Hz with a step size of 0.8 Hz; $\mathbf{f_s}$: from -8000 Hz to -588 Hz and 588 Hz to 8000 Hz with a step size of 4.9 Hz), respectively, $c_1$ (= -0.80838), $c_2$ (= 0.0028), and $c_3$ (= $10^{-5}$) are constants.

The alternative reward function of the $B_1$-insensitive selective inversion pulse (Supplementary Fig. 22) is set to have the same design specifications (the same mean $M_z$ over the target region (frequency range = -200 Hz to 200 Hz and $B_1$ range = 0.5 to 2.0)) as that of the HS pulse and the negligible magnitudes of $M_z$ outsides of the target region while minimizing ENG. The definition is as follows:

$$- \max(E[\mathbf{M_{z,DeepRF}}(\mathbf{B_1}, \mathbf{f_b})], c_1) - \max\left(\max_{\text{element}}(|1 - \mathbf{M_{z,DeepRF}}(\mathbf{B_1}, \mathbf{f_s})|) - c_2, 0\right)$$
$$- c_3 \cdot \text{ENG}_{\text{RF}}$$

(9)

where $\mathbf{M_{z,DeepRF}}(\mathbf{B_1}, \mathbf{f})$ is z-magnetizations over the target region, $\mathbf{B_1}$ is $B_1$ scaling factors from 0.5 to 2.0 with a step size of 0.03, $\mathbf{f_b}$ and $\mathbf{f_s}$ represent frequency ranges over the target region and outside of the target region ($\mathbf{f_b}$: from -200 Hz to 200 Hz with a step size of 4 Hz;

$f_s$: from -8000 Hz to -568 Hz and 568 Hz to 8000 Hz with a step size of 74.3 Hz), respectively, $c_1$ (= -0.906), $c_2$ (= 0.01), and $c_3$ (= $10^{-5}$) are constants.

**Conventional pulses**

For the conventional slice-selective pulses, SLR algorithm[13] implemented in MATPULSE (cind.ucsf.edu/education/software/matpulse)[47] and SigPy (github.com/jonbmartin/sigpy-rf)[48] was used. The input parameters for the excitation pulse were as follows: angle = 90°, time-step = 0.01 ms, number of time-steps = 256, duration = 2.56 ms, bandwidth = 2 kHz, maximum passband/stopband ripples = 1%, and the full width at half maximum of the slice profile was 2570 Hz. For the inversion pulse, the parameters were as follows: angle = 180°, time-step = 0.02 ms, number of time-steps = 256, duration = 5.12 ms, bandwidth = 500 Hz, maximum passband/stopband ripples = 1%, and the full width at half maximum of the slice profile was 836 Hz. In both pulses, the bandwidth convention was set as the maximum, and the filter option was tested for both least-squares using MATPULSE (default) and equi-ripple using SigPy. When designed, the maximum ripple conditions were slightly violated (1.5% maximum stopband ripple for excitation and 1.7% maximum passband ripple for inversion) as reported in the other studies[13,47].

For the conventional $B_1$-insensitive volume inversion pulse, we tested four adiabatic pulse designs: Hanning, Gauss, Lorentz, and HS pulses[31,32]. The target inversion ranges of $B_1$ and frequencies (i.e., target region) were from 0.5 to 2.0 and from -200 Hz to 200 Hz, respectively, which are common ranges at 3T[49] (time-step = 0.03125 ms, and number of time-steps = 256, duration = 8 ms). We aimed to optimize each adiabatic pulse to have the mean $M_z$ less than -0.9 over the target region with the smallest ENG. To achieve this aim, a grid search of the pulse parameters ($\omega_1^{max}$, $\beta$, and A)[32] was performed for the ranges of $\omega_1^{max}$ (from 2 mG to 200 mG), $\beta$ (from 100 rad/sec to 10000 rad/sec), and A (from 10 Hz to 2000 Hz) for each adiabatic pulse. After the search, an HS pulse with $\omega_1^{max}$ = 82 mG, $\beta$ = 3600 rad/sec, and A = 340 Hz reported the smallest ENG, and, therefore, was chosen as the final design.

In the case of the $B_1$-insensitive selective inversion pulse, the optimized HS pulse above was set as the reference pulse, and a grid search was performed for the four adiabatic pulse designs to find a better RF pulse that satisfies the following criteria: the mean $M_z$ was less than -0.9 in the target region, and both ENG and BW (full width at half maximum) at $B_1$ = 1 were smaller than those of the reference pulse. The search found no RF better than the reference pulse, so the reference pulse was chosen as the final design.

**MRI experiments**

MRI scans were performed at a 3T MRI system (Siemens MAGNETOM Trio) using a body coil for transmission and a 32-channel head coil for reception. A cylindrical water phantom was scanned. To obtain the experimental slice profile of the DeepRF or SLR slice-selective excitation pulse, a sagittal plane image was acquired after an axial excitation. Then, the central 100 lines of the sagittal image were averaged to generate the slice profile. The scan parameters were as follows: field of view = 256×256 mm$^2$, resolution = 0.5×0.5 mm$^2$, echo time (TE) = 4.6 ms, repetition time = 500 ms, and bandwidth = 540 Hz/Px. For the inversion profile of the DeepRF or SLR slice-selective inversion pulse, an axial inversion was followed by an adiabatic half passage pulse[32] (AHP; angle = 482° and duration = 4.5 ms) when the phantom signal is nulled (inversion time = 69.4 ms; phantom $T_1$ = 100 ms). After that, the inversion profile was created by averaging the central 100 lines of the sagittal image. The scan parameters were the same as the excitation pulse experiment except for TE = 7.8 ms. In the case of the $B_1$-insensitive volume inversion or selective inversion pulses, a two-dimensional inversion profile (i.e., $M_z$ over ranges of $B_1$ and frequencies) was acquired by obtaining multiple images with scaled $B_1$ amplitudes from 0.5 to 2.0 with a step size of 0.1 for the inversion pulse. A sagittal image was acquired after axial inversion and AHP excitation same as the slice-selective inversion pulse experiment. The inversion profile was created by filling each row (i.e., $M_z$ over frequencies for fixed $B_1$) with the mean of the central 100 lines of each image. In all scans, the receive coil sensitivity was corrected by dividing the image by a reference image, which was obtained from a reference scan (gradient-echo; excitation RF = AHP; no inversion pulse; all other parameters the same as above including the body coil for transmission and the head coil for reception).

**Other information**

DeepRF was implemented by modifying the Python codes of Niraj Amalkanti (github.com/namalkanti/bloch-simulator-python) using PyTorch[45]. The computing environment was Intel® Xeon® Gold 5218 2.30 GHz CPU, NVIDIA Quadro RTX 8000 GPU, and 128 GB memory.

## DATA AVAILABILITY

All processed data shown in the figures and table are available at https://github.com/SNU-LIST/DeepRF/tree/master/data.

## CODE AVAILABILITY

The source code of DeepRF is available at https://github.com/SNU-LIST/DeepRF[50].

## ACKNOWLEDGEMENT

This work was supported by the National Research Foundation of Korea (NRF-2021R1A2B5B03002783), Samsung Research Funding & Incubation Center of Samsung Electronics (SRFC-IT1801-09), RadiSen, INMC, and IOER at Seoul National University.

## AUTHOR CONTRIBUTIONS

D. S. conceived the study, conducted the experiments, and wrote the paper together with J. L., while Y. K. implemented the algorithm. C. O. and J. L. assisted with the experimental data interpretation. H. A. helped with the problem formulation. J. P. and J. K. contributed to the development of the concept of the study. All authors reviewed and commented on the manuscript.

## COMPETING INTERESTS

The authors declare the following competing interests: two patents are disclosed (in the order of patent applicant, names of inventors, application number, status of application, specific aspect of manuscript covered in patent application).

1. Seoul National University, D. S. and J. L., US Patent No. 17/168,274, patent pending, **RF refinement module** in **METHODS**

2. Seoul National University, D. S. and J. L., Korea Patent No. 10-2020-0106569, patent pending, **RF refinement module** in **METHODS**

| | RF energy reduction (%) | Mean magnitude or $M_z$ over BW | Maximum passband ripple (%) | Maximum stopband ripple (%) | Mean $M_z$ |
|---|---|---|---|---|---|
| Slice-selective excitation | 15.6 ± 1.2 | 0.93 ± 0.00 | 2.3 ± 0.2 | 1.5 ± 0.0 | - |
| Slice-selective inversion | 10.8 ± 0.0 | -0.81 ± 0.00 | 1.9 ± 0.0 | 0.4 ± 0.0 | - |
| $B_1$-insensitive volume inversion | 9.3 ± 0.0 | - | - | - | -0.91 ± 0.00 |
| $B_1$-insensitive selective inversion | 1.6 ± 0.0 | - | - | - | -0.90 ± 0.00 |

**Table 1.** Summary of the reproducibility results of the four DeepRF pulses. All values are given as mean ± standard deviation.

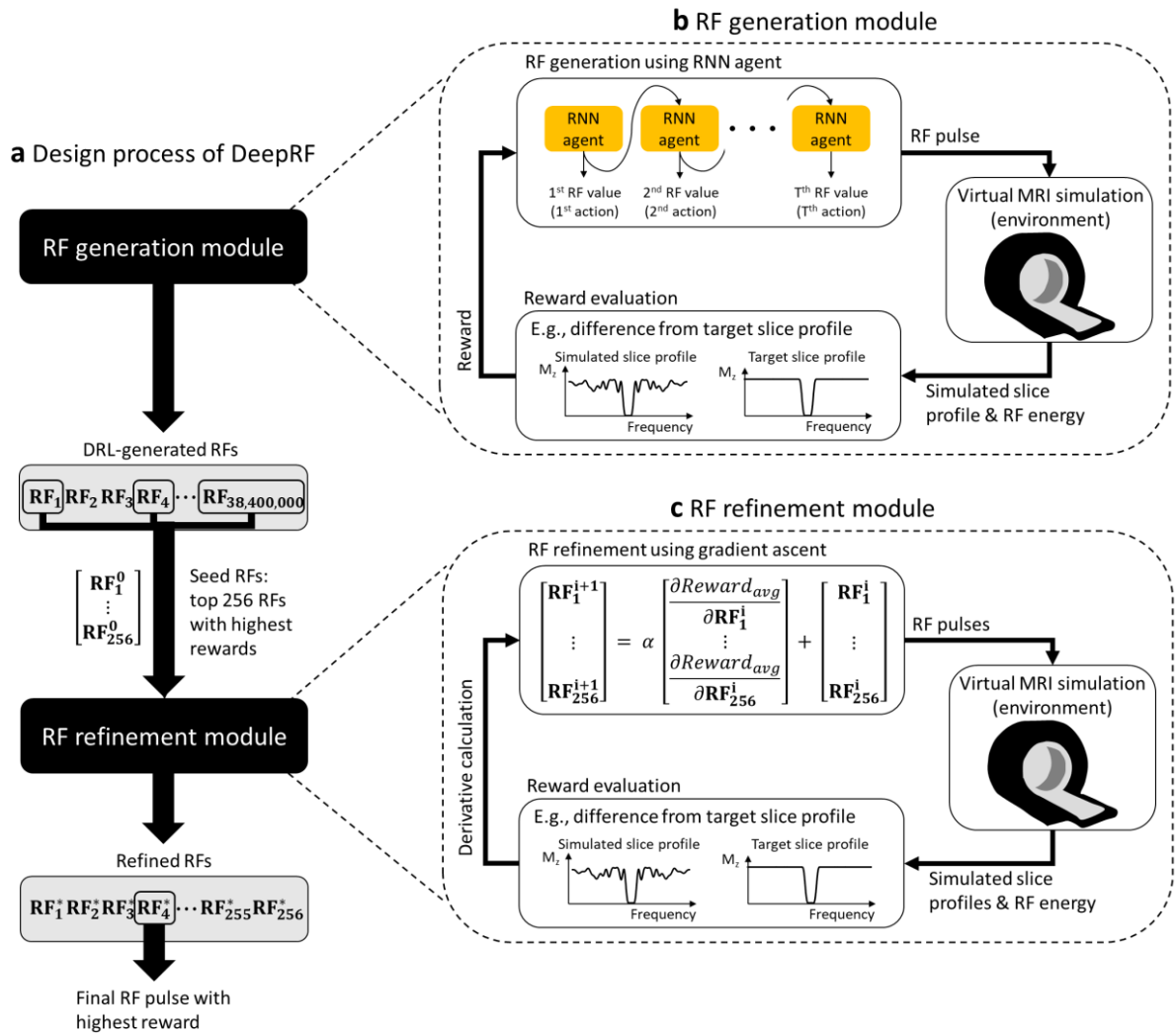

**Figure 1**. The design process of DeepRF. (a) In DeepRF, an RF pulse is created by a sequence of an RF generation module and an RF refinement module. In the RF generation module, deep reinforcement learning is utilized to produce a large number of RF pulses from which the top 256 pulses with the highest rewards are chosen as seed RF pulses. These seed RF pulses are further improved via gradient ascent in the RF refinement module. Finally, the RF pulse with the highest reward is chosen as the final design. (b) The RF generation module utilizes a recurrent neural network (RNN) agent to produce an RF pulse, which is fed into a virtual MRI simulation for a slice/inversion profile and RF energy. Then, these results are evaluated, calculating a reward which is fed back to the RNN agent, updating the weights of RNN for improved RF pulse generation. (c) In the RF refinement module, the seed pulses from the RF generation module are optimized via gradient ascent by iteratively updating RF pulses using gradient, which had a similar loop as the generation module.

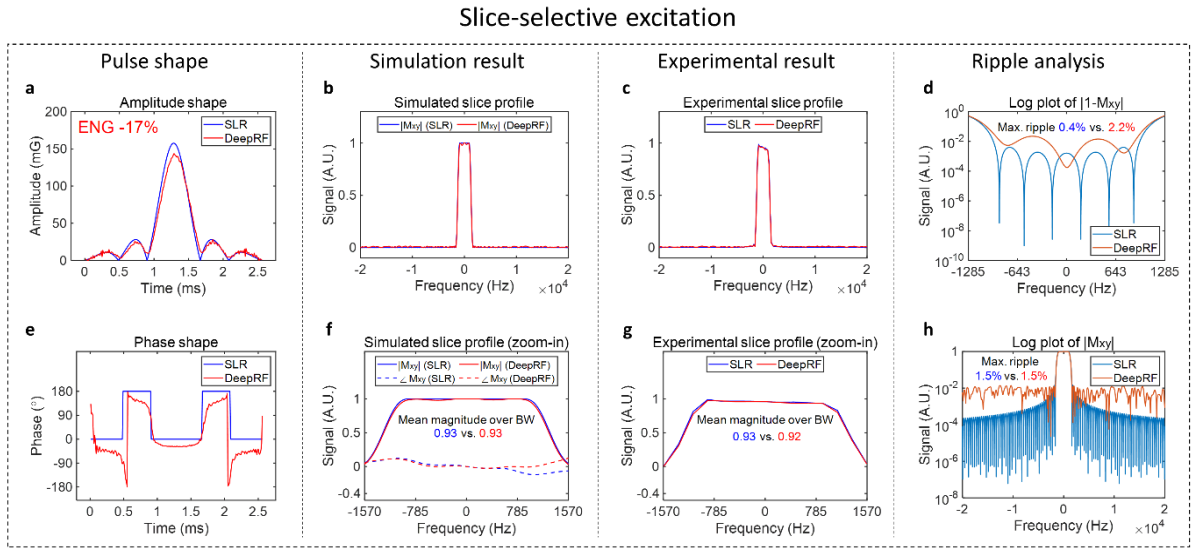
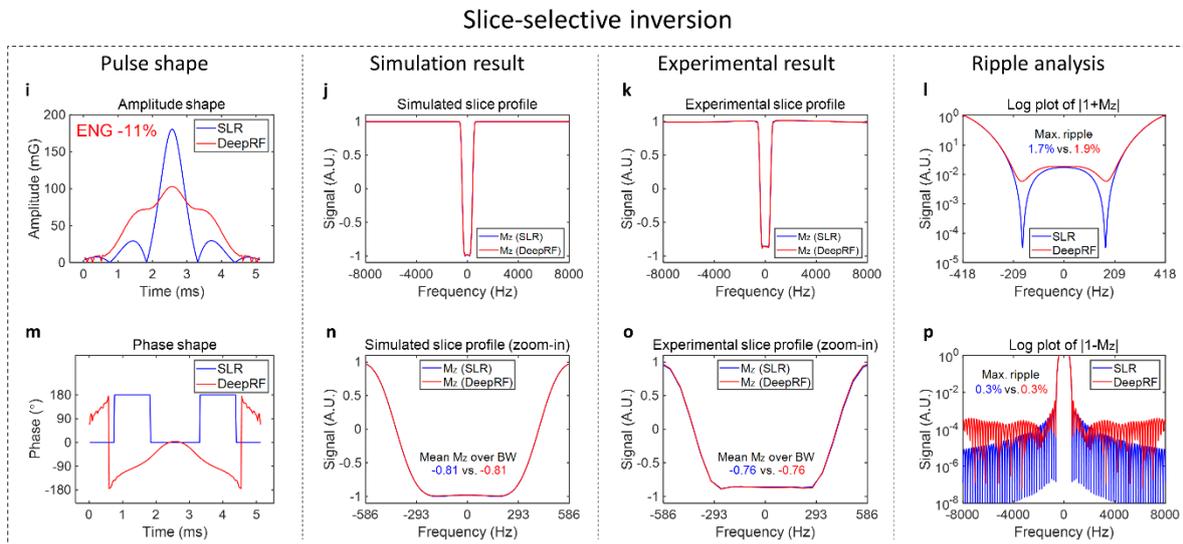

**Figure 2.** The results of the slice-selective excitation pulses (a-h) and inversion pulses (i-p). In the slice-selective excitation pulses, the amplitude (a) and phase (b) of the DeepRF pulse mimic those of the SLR pulse, while reporting 17% smaller RF ENG. The simulated slice profiles of the two pulses match well (b; zoomed-in figure in f), and the experimental slice profiles are also similar (c; zoomed-in figure in g). A detailed analysis of the ripples is shown for passband (d) and stopband (h). In the slice-selective inversion pulses, the amplitude (i) and phase (m) of the DeepRF pulse are substantially different from those of the SLR pulse, reporting 11% smaller RF ENG. The simulated slice profiles are highly comparable (j; zoomed-in figure in n), and the experimental slice profiles show a good match between the two pulses (k; zoomed-in figure in o). A ripple analysis is shown for passband (l) and stopband (p). Comparison results between equi-ripple-designed SLR pulses and DeepRF pulses are in Supplementary Fig. 1 for excitation and Supplementary Fig. 3 for inversion.

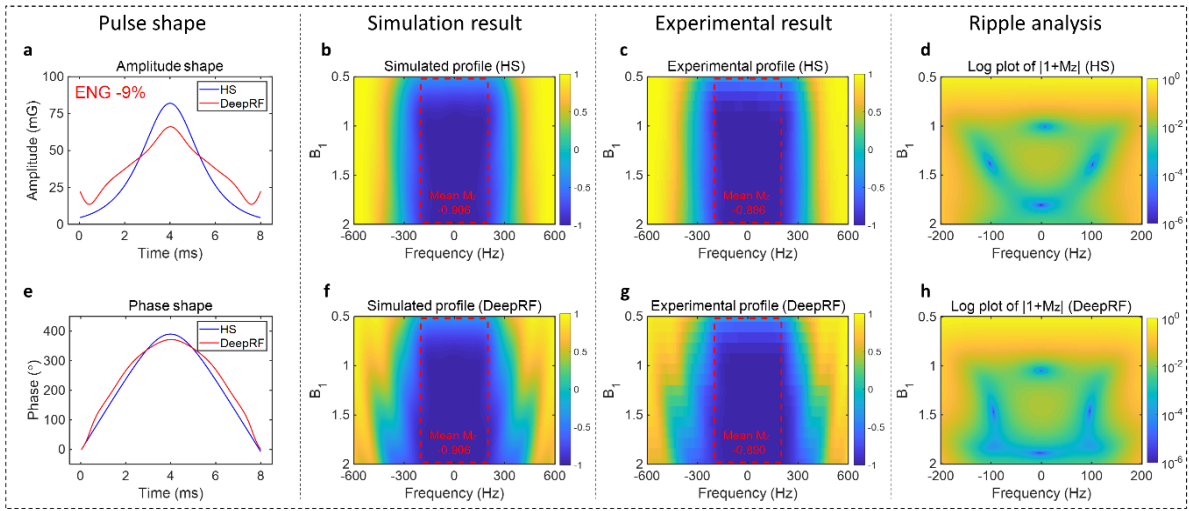

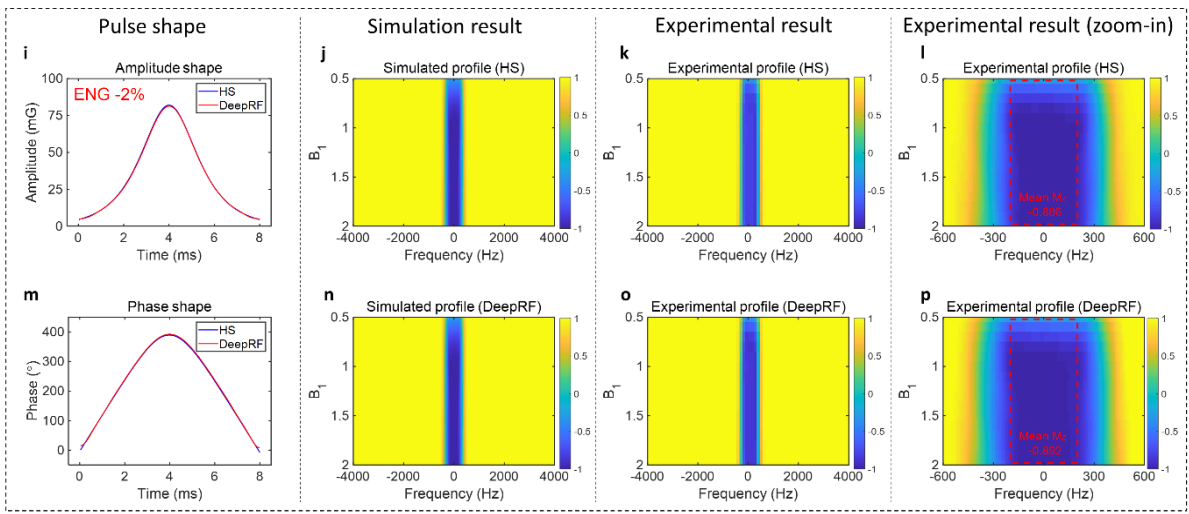

**Figure 3.** The results of the B$_1$-insensitive volume inversion pulses (a-h) and selective inversion pulses (i-p). In the B$_1$-insensitive volume inversion pulses, the amplitude of the DeepRF pulse is substantially different from that of the HS pulse, reporting 9% smaller RF ENG (a: amplitude; b: phase). Over the target region, the simulated inversion profiles show a good match between the two pulses (b vs. f; see red dotted boxes for the target region), but their details are different (d vs. h). The experimental profiles agree with the simulation results (b vs. c; f vs. g). For the B$_1$-insensitive selective inversion pulses, the amplitude (i) and phase (m) of the two pulses are surprisingly similar to each other despite the fact that no knowledge of any conventional RF pulses is trained in DeepRF. The simulated profiles of the two pulses are well-matched (j vs. n), and the experimental profiles agree well with the simulated profiles (j vs. k; n vs. o; see l and p for zoomed-in images).

## Analysis of slice-selective inversion pulse

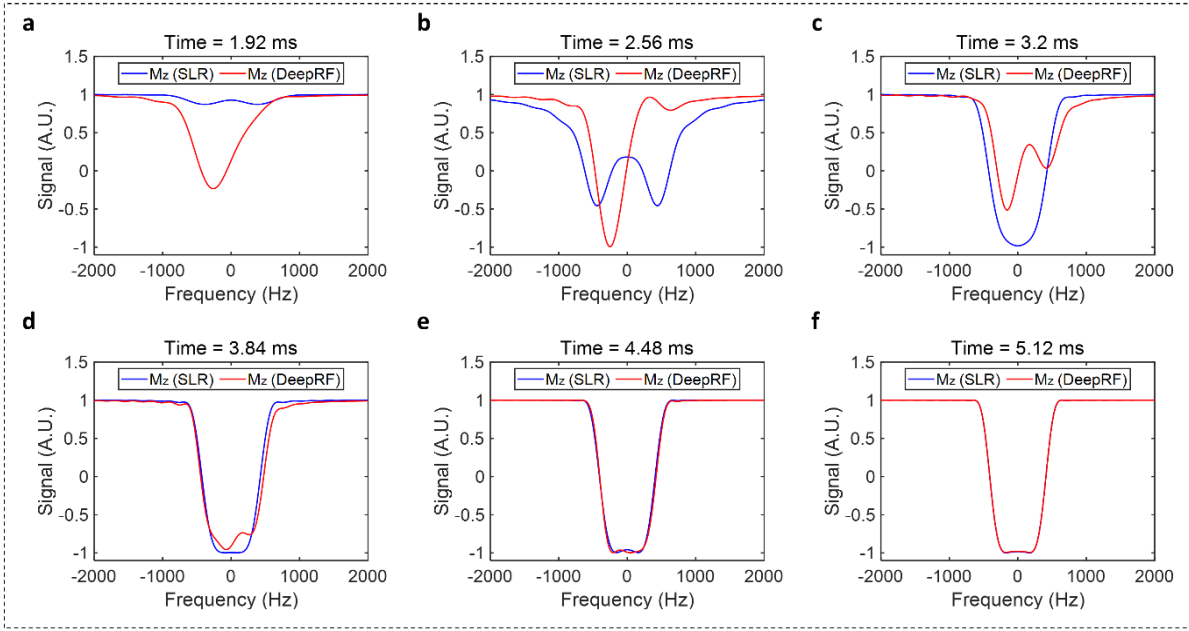

## Analysis of $B_1$-insensitive volume inversion pulse

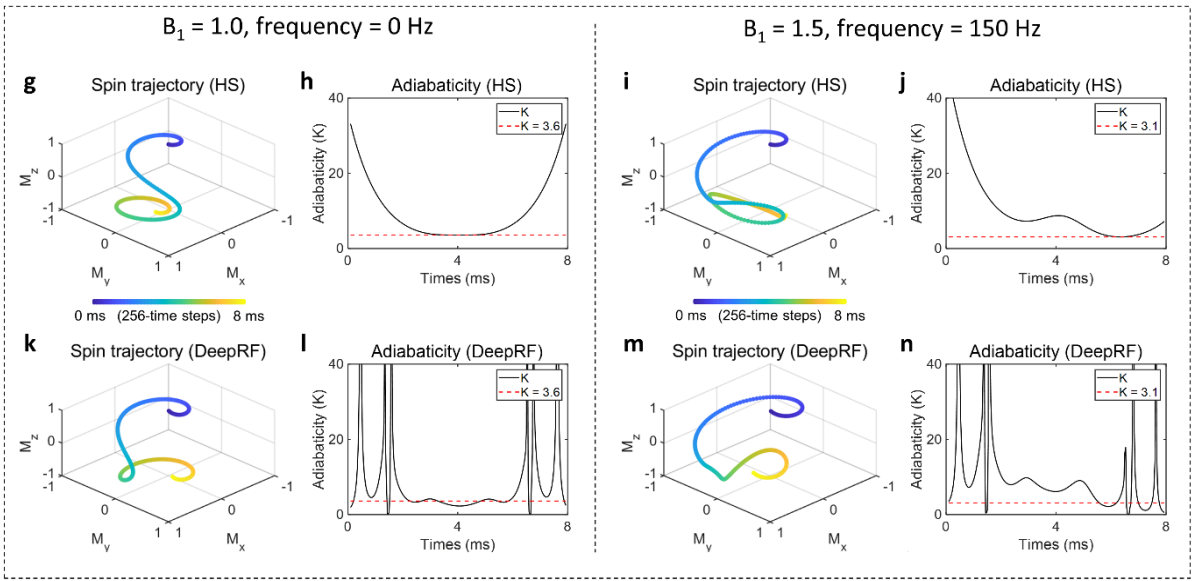

**Figure 4.** The analysis results of the DeepRF pulses. The snapshots of the slice profile of the slice-selective inversion pulses are shown at six-time points, revealing substantially different inversion processes in the two pulses (a-f). The spin trajectories and adiabaticity of the $B_1$-insensitive volume inversion pulse are plotted for the two conditions of $B_1$ and off-resonance frequencies ($B_1$ = 1.0 and frequency = 0 Hz; $B_1$ = 1.5 and frequency = 150 Hz) (g-n). The DeepRF pulse reveals substantially different spin trajectories from the HS pulse in the two conditions (g vs. k; i vs. m). In the HS pulse, the adiabatic condition (adiabaticity >> 1) are well-satisfied throughout the pulse duration (h, j). On the other

hand, the adiabaticity of the DeepRF pulse is often smaller than the minimum values of the HS pulse (l, n; see red dotted lines), reporting the minimum adiabaticity of 0.1 (l) and 0.04 (n). This result indicates that DeepRF violates the adiabatic condition and has a different inversion mechanism from the conventional design principle.

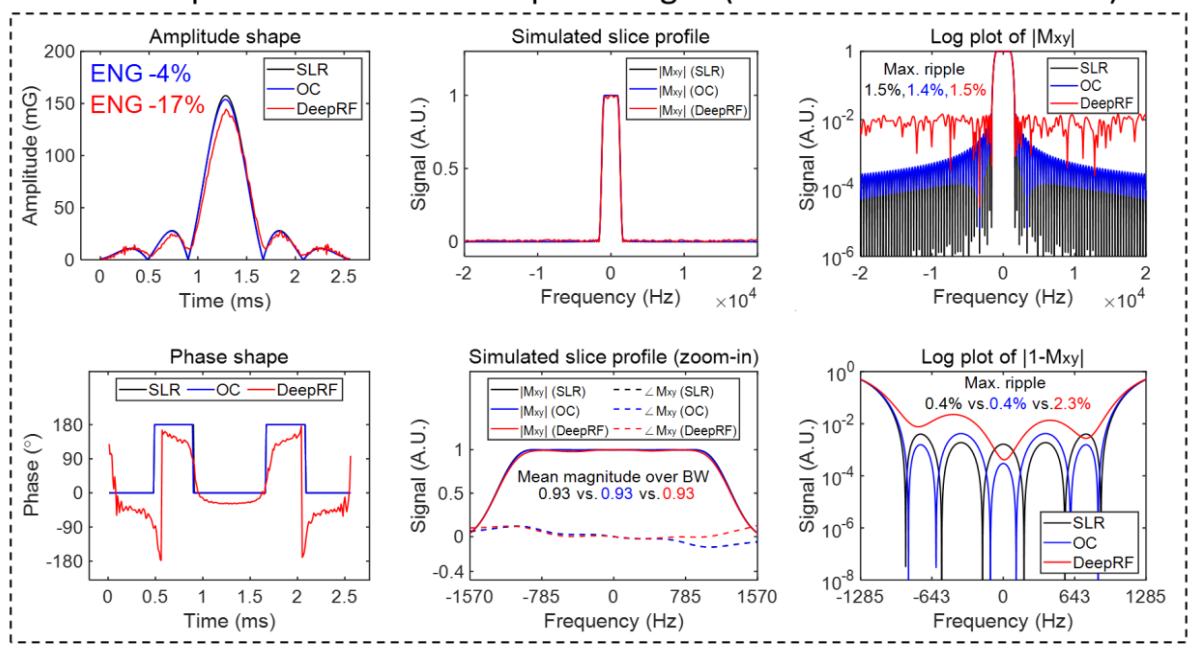

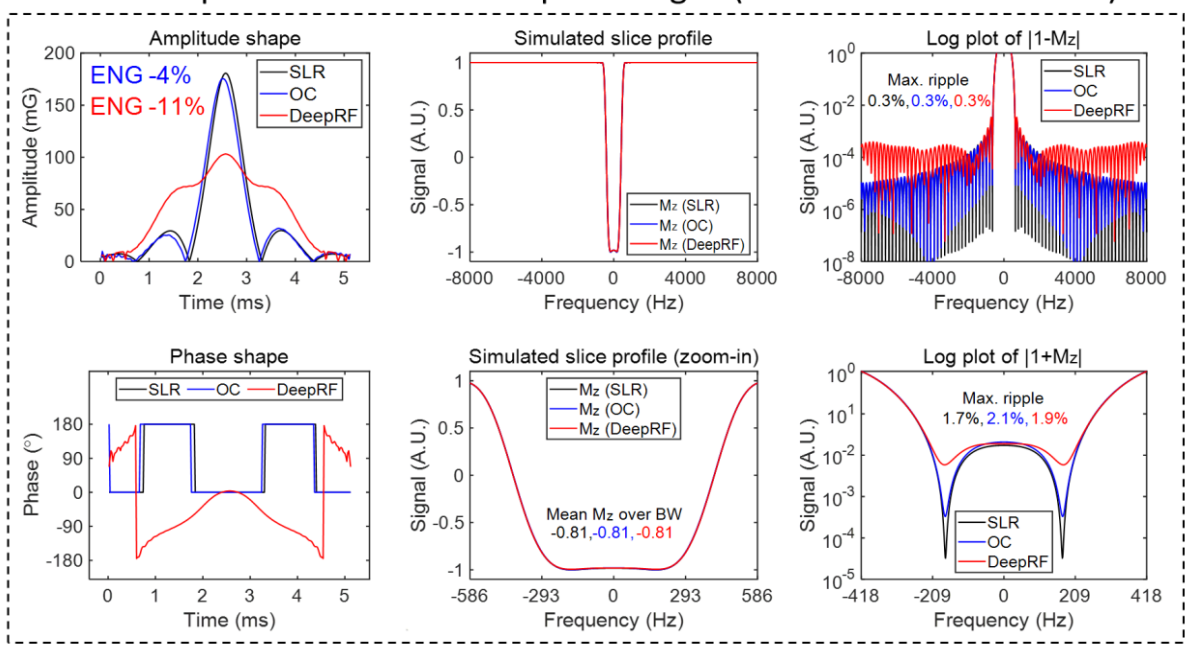

**Figure 5.** The comparison results between the SLR, OC, and DeepRF pulses in the slice-selective excitation and inversion RF designs. The DeepRF pulses show larger ENG reduction than the OC pulses when compared with those of the SLR pulses (OC: -4% for excitation and -4% for inversion; DeepRF: -17% for excitation and -11% for inversion).

# REFERENCES

1	Silver, D. *et al.* Mastering the game of Go without human knowledge. *Nature* **550**, 354-359, doi:10.1038/nature24270 (2017).

2	Wu, Y. *et al.* Google's neural machine translation system: Bridging the gap between human and machine translation. Preprint at https://arxiv.org/abs/1609.08144 (2016).

3	Yu, J. *et al.* Generative image inpainting with contextual attention. *Proceedings of the IEEE Conference on Computer Vision and Pattern Recognition*, 5505-5514 (2018).

4	Segler, M. H., Preuss, M. & Waller, M. P. Planning chemical syntheses with deep neural networks and symbolic AI. *Nature* **555**, 604-610 (2018).

5	Runge, F., Stoll, D., Falkner, S. & Hutter, F. Learning to design RNA. *Proceeding of the International Conference on Learning Representations* (2019).

6	Wang, H., Yang, J., Lee, H.-S. & Han, S. Learning to design circuits. *Proceeding of the Conference on Neural Information Processing Systems* (2018).

7	Popova, M., Isayev, O. & Tropsha, A. Deep reinforcement learning for de novo drug design. *Science Advances* **4** (2018).

8	Sutton, R. S. & Barto, A. G. *Reinforcement Learning: An Introduction*.   (MIT Press, Cambridge, MA, USA, 2018).

9	Bell, M. R. Information theory and radar waveform design. *IEEE Transactions on Information Theory* **39**, 1578-1597 (1993).

10	Simpson, D. H., Chin, C. T. & Burns, P. N. Pulse inversion Doppler: a new method for detecting nonlinear echoes from microbubble contrast agents. *IEEE Transactions on Ultrasonics, Ferroelectrics, and Frequency Control* **46**, 372-382 (1999).

11	Tarantola, A. *Inverse problem theory and methods for model parameter estimation*.

(SIAM, Philadelphia, PA, USA, 2005).

12    Bloch, F. Nuclear induction. *Physical Review* **70**, 460 (1946).

13    Pauly, J., Le Roux, P., Nishimura, D. & Macovski, A. Parameter relations for the Shinnar-Le Roux selective excitation pulse design algorithm. *IEEE Transactions on Medical Imaging* **10**, 53-65 (1991).

14    Abragam, A. *The principles of nuclear magnetism*.    (Oxford university press, 1961).

15    Conolly, S., Nishimura, D. & Macovski, A. Optimal control solutions to the magnetic resonance selective excitation problem. *IEEE Transactions on Medical Imaging* **5**, 106-115 (1986).

16    Khaneja, N., Reiss, T., Kehlet, C., Schulte-Herbrüggen, T. & Glaser, S. J. Optimal control of coupled spin dynamics: design of NMR pulse sequences by gradient ascent algorithms. *Journal of Magnetic Resonance* **172**, 296-305 (2005).

17    Rosenfeld, D. & Zur, Y. Design of adiabatic selective pulses using optimal control theory. *Magnetic Resonance in Medicine* **36**, 401-409 (1996).

18    Rund, A., Aigner, C. S., Kunisch, K. & Stollberger, R. Magnetic resonance RF pulse design by optimal control with physical constraints. *IEEE Transactions on Medical Imaging* **37**, 461-472 (2017).

19    Rund, A., Aigner, C. S., Kunisch, K. & Stollberger, R. Simultaneous multislice refocusing via time optimal control. *Magnetic Resonance in Medicine* **80**, 1416-1428 (2018).

20    Xu, D., King, K. F., Zhu, Y., McKinnon, G. C. & Liang, Z. P. Designing multichannel, multidimensional, arbitrary flip angle RF pulses using an optimal control approach. *Magnetic Resonance in Medicine* **59**, 547-560 (2008).

21    Vinding, M. S., Maximov, I. I., Tošner, Z. & Nielsen, N. C. Fast numerical design of spatial-selective rf pulses in MRI using Krotov and quasi-Newton based optimal control methods. *The Journal of Chemical Physics* **137**, 054203 (2012).

22   Vinding, M. S., Guérin, B., Vosegaard, T. & Nielsen, N. C. Local SAR, global SAR, and power-constrained large-flip-angle pulses with optimal control and virtual observation points. *Magnetic Resonance in Medicine* **77**, 374-384 (2017).

23   Loecher, M., Magrath, P., Aliotta, E. & Ennis, D. B. Time-optimized 4D phase contrast MRI with real-time convex optimization of gradient waveforms and fast excitation methods. *Magnetic Resonance in Medicine* **82**, 213-224 (2019).

24   Shang, H. *et al.* Multiband RF pulses with improved performance via convex optimization. *Journal of Magnetic Resonance* **262**, 81-90 (2016).

25   Vinding, M. S., Aigner, C. S., Schmitter, S. & Lund, T. E. DeepControl: 2D RF pulses facilitating B1+ inhomogeneity and B0 off-resonance compensation in vivo at 7T. *Magnetic Resonance in Medicine*, 3308-3317 (2021).

26   Vinding, M. S., Skyum, B., Sangill, R. & Lund, T. E. Ultrafast (milliseconds), multidimensional RF pulse design with deep learning. *Magnetic Resonance in Medicine* **82**, 586-599 (2019).

27   Mirfin, C., Glover, P. & Bowtell, R. Optimisation of parallel transmission radiofrequency pulses using neural networks. *Proceeding of the 26th Annual Meeting of ISMRM* (2018).

28   Zhang, Y. *et al.* Multi-task convolutional neural network-based design of radio frequency pulse and the accompanying gradients for magnetic resonance imaging. *NMR in Biomedicine* **34** (2021).

29   Goodfellow, I., Bengio, Y., Courville, A. & Bengio, Y. *Deep learning*. (MIT Press, Cambridge, MA, USA, 2016).

30  Silver, M. S., Joseph, R. & Hoult, D. Highly selective π2 and π pulse generation. *Journal of Magnetic Resonance* **59**, 347-351 (1984).

31  Garwood, M. & DelaBarre, L. The return of the frequency sweep: designing adiabatic pulses for contemporary NMR. *Journal of magnetic resonance* **153**, 155-177 (2001).

32  Tannús, A. & Garwood, M. Adiabatic pulses. *NMR in Biomedicine* **10**, 423-434 (1997).

33  Henderson, P. *et al.* Deep reinforcement learning that matters. *Proceeding of the AAAI Conference on Artificial Intelligence* (2018).

34  Glorot, X. & Bengio, Y. Understanding the difficulty of training deep feedforward neural networks. *Proceedings of the International Conference on Artificial Intelligence and Statistics*, 249-256 (2010).

35  Vinyals, O. *et al.* Grandmaster level in StarCraft II using multi-agent reinforcement learning. *Nature* **575**, 350-354, doi:10.1038/s41586-019-1724-z (2019).

36  Zhu, B., Liu, J. Z., Koonjoo, N., Rose, B. R. & Rosen, M. S. Automated pulse sequence generation (AUTOSEQ) using Bayesian reinforcement learning in an MRI physics simulation environment. *Proceeding of the 26th Annual Meeting of ISMRM* (2018).

37  Walker-Samuel, S. Using deep reinforcement learning to actively, adaptively and autonomously control of a simulated MRI scanner. *Proceeding of the 27th Annual Meeting of ISMRM* (2019).

38  David Y Zeng, Christopher M Sandino, Dwight G Nishimura, Shreyas S Vasanawala & Cheng, J. Y. Reinforcement learning for online undersampling pattern optimization. *Proceeding of the 27th Annual Meeting of ISMRM* (2019).

39  Pineda, L., Basu, S., Romero, A., Calandra, R. & Drozdzal, M. Active MR k-space sampling with reinforcement learning. *Proceeding of the International Conference on*

*Medical Image Computing and Computer-Assisted Intervention* (2020).

40  Bahadir, C. D., Wang, A. Q., Dalca, A. V. & Sabuncu, M. R. Deep-learning-based optimization of the under-sampling pattern in MRI. *IEEE Transactions on Computational Imaging* **6**, 1139-1152 (2020).

41  Meyer, C. H., Pauly, J. M., Macovskiand, A. & Nishimura, D. G. Simultaneous spatial and spectral selective excitation. *Magnetic Resonance in Medicine* **15**, 287-304 (1990).

42  Yip, C. y., Fessler, J. A. & Noll, D. C. Iterative RF pulse design for multidimensional, small-tip-angle selective excitation. *Magnetic Resonance in Medicine* **54**, 908-917 (2005).

43  Cho, K. *et al.* Learning phrase representations using RNN encoder-decoder for statistical machine translation. *Proceeding of the Conference on Empirical Methods in Natural Language Processing* (2014).

44  Schulman, J., Wolski, F., Dhariwal, P., Radford, A. & Klimov, O. Proximal policy optimization algorithms. Preprint at https://arxiv.org/abs/1707.06347 (2017).

45  Paszke, A. *et al.* Automatic differentiation in pytorch. *Proceeding of the Conference on Neural Information Processing Systems* (2017).

46  Kingma, D. P. & Ba, J. Adam: A Method for Stochastic Optimization. *Proceeding of the International Conference for Learning Representations* (2015).

47  Matson, G. B. An integrated program for amplitude-modulated RF pulse generation and re-mapping with shaped gradients. *Magnetic resonance imaging* **12**, 1205-1225 (1994).

48  Martin, J. B. *et al.* SigPy.RF: Comprehensive Open-Source RF Pulse Design Tools for Reproducible Research. *Proceedings of the International Society of Magnetic Resonance in Medicine, Montréal, QC* **4819** (2020).

49  Stockmann, J. P. *et al.* A 32-channel combined RF and B0 shim array for 3T brain imaging. *Magnetic Resonance in Medicine* **75**, 441-451 (2016).

50      Shin, D. *et al.* DeepRF: (v1.0). Zenodo. https://doi.org/10.5281/zenodo.5529394 (2021).